\documentstyle[11pt,aaspp4]{article}

\begin{document}
\title{Do Proto-Jovian Planets Drive Outflows?}

\author{
A.~C.\ Quillen\altaffilmark{1}  \&
D. E. Trilling\altaffilmark{2}
}
\altaffiltext{1}{Steward Observatory, The University of Arizona, Tucson, AZ 85721;
 aquillen@as.arizona.edu}
\altaffiltext{2}{Lunar and Planetary Laboratory, The University of Arizona, Tucson, AZ 85721; trilling@lpl.arizona.edu}

\begin{abstract}
We discuss the possibility that gaseous giant planets 
drive strong outflows during early phases of their formation.  
We consider the range of parameters appropriate for
magneto-centrifugally driven stellar and disk outflow models 
and find that if the proto-Jovian planet or accretion disk had a magnetic
field of $\gtrsim 10$ Gauss
and moderate mass inflow accretion rates through the disk of 
less than $\sim 10^{-7} M_J$/yr
that it is possible to drive an outflow.
Estimates based both on scaling from empirical laws observed
in proto-stellar outflows and the magneto-centrifugal
disk and stellar+disk wind models 
suggest that winds with mass outflow rates of order $10^{-8} M_J$/yr
and velocities of order $\sim 20$ km/s could be driven from 
proto-Jovian planets.  
Prospects for detection and some implications for the formation 
of the solar system are briefly discussed.
\end{abstract}


\section {Introduction}

Outflows and jets are ubiquitous in astronomy, and are observed 
in a wide range of astrophysical objects ranging on the
largest and most energetic scales from active galactic nuclei, 
to X-ray binaries and protostars on the smallest scales.
One property that these objects have in common is that they are
suspected to be undergoing accretion from a gaseous disk.
It is possible that outflows play a crucial role in removing
angular momentum from an accretion disk (e.g.\ \cite{shu87}) and so
are a natural consequence of accretion.

The prograde, low eccentricity, equatorial, low inclination orbits of the
regular satellites of the giant planets of our solar system 
suggest that these planets had 
disks during some phase of their formation.
Additionally, the Galilean satellites have
densities which decrease with increasing
distance from Jupiter (Io, Europa, Ganymede,
and Callisto, from innermost to outermost).
If the satellites formed from a gaseous disk 
then this decrease in density,  
due to an increase in volatile contents,
suggests that the temperature of the circum-Jovian disk
varied with radius.  The early circum-Jovian
environment can be viewed as analogous in many ways
to a forming planetary system (\cite{stevenson86}). 
However we note that formation models for the Galilean satellites
are varied and not necessarily straightforward
(e.g.\ \cite{lunine}; \cite{anderson}; \cite{mck97}; \cite{coradini};
\cite{pri81}; \cite{pri93}; \cite{kor91}; \cite{kor90}; \cite{pol90}).  
Fluid simulations of gaseous planet formation 
can also show such an accretion disk.
Because Jovian planets have strong magnetic fields 
and could have had accretion disks, 
we are prompted to consider the possibility that they too produced
magnetically accelerated outflows during some early phase of their formation.

In proto-stellar objects,
observations show that the mechanical energy in 
thrust carried in molecular outflows 
can be large compared 
with the bolometric luminosity of the central young stellar object 
(\cite{lad85}).
As a result the acceleration mechanism cannot rely on radiation 
pressure alone, and other forces are required to drive the outflow.  
The notion that rapid rotation coupled with
strong magnetic fields could drive outflows was 
explored in the context of the solar wind by \cite{mes68}.
Subsequently \cite{bla82} introduced an elegant scale free
hydromagnetic wind model for jet acceleration from a magnetized
disk which gives centrifugally driven wind solutions.  These 
solutions also have the possibility of being self-collimated 
(\cite{hey89}).  
Similar outflow models were applied directly to proto-stellar objects 
(\cite{pud83}, \cite{kon89}, \cite{pel92}, \cite{har82} and others).  
One requirement of the disk magneto-centrifugal wind models is that
the field lines bend outwards from the disk by more than 
$30^\circ$ from the normal to the disk.
However it is unclear how the internal disk structure 
maintains this field geometry
(\cite{war93}, see the review of \cite{shu93}).
This problem provoked a modification of the stellar and disk wind 
accelerated model 
of \cite{shu88}, commonly referred to as the X-wind model  
(\cite{shu94a} and subsequent papers; \cite{shu94b}, \cite{shu95}),
which considers the 
the truncation of the accretion disk from the magnetosphere
of the spinning central protostar.

The mechanical luminosity and force or thrust (rate of momentum
injection) of bipolar molecular outflows scales with the radiant
luminosity of the driving stellar source (\cite{lad85}, \cite{cab92}).
The possibility that low mass (brown dwarf or lower mass) 
companions in binary systems might drive outflows
was suggested by \cite{wol90} who scaled from properties observed
empirically in proto-stellar outflows and argued that brown
dwarfs would be capable of driving outflows.  

In this section we discuss why or why not
we might expect the accretion physics of Jovian (or giant) planets
to be similar to that of low mass protostars.
In \S 2 we estimate ranges for the magnetic field, accretion rate in
the disk and radius for which magneto-centrifugal outflow models could operate.
In \S 3 with a similar approach to \cite{wol90}
we compare outflow velocities and mass outflow rates 
consistent with the empirical scaling laws to those appropriate for the
theoretical models.

\subsection{Why might the physics of accretion be {\it similar}
for protostars and proto-Jovian planets?}

We first mention the relevance of the magnetic field under physical conditions
appropriate to proto-stellar and proto-Jovian disks.
As is true in the inner radii of proto-stellar disks 
(e.g.\ \cite{gam96}, \cite{shu94a}), 
within a few planetary radii the proto-Jovian disk
is expected to be above 1000K (\cite{lunine})
and so thermal ionization 
results in a sufficiently high ionization fraction that the gas and 
magnetic field are well coupled (\cite{takata}).
Below a temperature of 1000K thermal ionization
does not dominate.  However ionization caused by cosmic rays, 
live radioactive nuclides and X-ray emission
from the central star or planet can produce enough ionization 
to couple the gas and magnetic field (\cite{lev91};
\cite{mor93}), but only in the outer layers of
a dense gas disk (\cite{gam96}; \cite{takata}; \cite{gla97}).  
As proposed by \cite{takata_b}
it is therefore likely that the dynamics of Jovian planets 
and their gaseous disks were influenced by the magnetic field 
as is true in the case of proto-stars.

The thrust of proto-stellar bipolar outflows is large enough that the stellar
radiative luminosity cannot account for the observed momentum and energy 
outflow.  A magnetic field is therefore expected to drive the acceleration of
the outflow, and the bolometric luminosity is not key.  
Protostellar molecular outflows have been identified
surrounding sources with luminosities as low as $0.2 L_\odot$
and higher than $10^6 L_\odot$ (\cite{bal91}).
The lowest luminosity sources are still capable of driving outflows
and may represent a new class of sources (\cite{and93}).
This enormous range of luminosities suggests that the underlying physics
of accretion and outflow is not primarily dependent on radiative luminosity.  

The early phase of contraction during which the luminosity depends on 
gravitational energy and not nuclear burning (nucleosynthesis), 
should be similar for low mass stars and Jovian planets.
The spectral energy distribution of Class I (embedded) 
sources have been successfully
modeled as embryonic stellar cores surrounded by circumstellar disks and 
massive gas and dust envelopes whose density structure is consistent
with rotating infalling molecular cloud cores (\cite{ada87}).
The bipolar outflow phase for protostars is 
likely to occur during during this early stage of collapse (or perhaps 
even earlier in the Class 0 sources).  
Between 
50\% and 80\% of luminous Class I sources show molecular outflows
suggesting that most young stellar objects go through an outflow stage
(\cite{mor91}) and that Class I sources spend a significant fraction
of their lifetime undergoing outflow (\cite{lad85}).
Subsequent studies find that {\it all} embedded proto-stellar
sources have some degree of outflow activity 
suggesting that the outflow phase and the infall/accretion phase coincide
(\cite{bon96}).

Though Jupiter is rotating quite quickly, it is currently 
spinning well below the centrifugal breakup speed.  Because
it has contracted since formation, its spin must have been even lower 
in the past.  Currently Jupiter's spin down 
is dominated by the magnetic breaking resulting from the
interaction of the solar wind and Jupiter's magnetosphere
(e.g.\ \cite{isb84}).  This torque is so low as to suggest that
Jupiter's angular momentum has
not changed significantly since formation.  However, since the sun 
was probably 
more active in the past, the spin down rate from this
process could have been more efficient when the solar system was younger.
If Jupiter was formed spinning near breakup
a remarkably efficient spin down process is required to account
for its present angular rotation rate.
Protostars are also observed to be spinning well below breakup speed.  
As proposed by \cite{takata_b} there may have been
a connection between the
angular momentum of the proto-Jovian disk and the planet,
as has been proposed for protostars (\cite{kon91}). 
\cite{takata_b}  have proposed that magnetic
interactions between the planet and the proto-Jovian disk
could provide sufficient torque to despin the planet
from rotational break-up to its present rotational state.
Another possibility is that angular momentum transfer
between the core and the outer envelope resulted in the ejection
of a high angular momentum disk (possibly forming the proto-Jovian
disk) and a relatively slowly spinning planet (\cite{kor91}).

\subsection{Why might the physics of accretion be {\it different}
for protostars and proto-Jovian planets?}

Formation scenarios of Jovian planets fall into two broad categories:
a) as a direct condensation from the proto-solar nebula (\cite{boss97}), and
b) in a two phase process where a rocky core is formed, and 
a gaseous envelope is subsequently accreted to form the planet's atmosphere
(e.g.\ \cite{bp86}; \cite{pod93}; \cite{pol96}; \cite{lissauer}).
The direct condensation scenario predicts near solar abundances 
whereas in the two phase scenarios the details of the model determine 
the core to envelope mass fractions (e.g.\ \cite{pol96} and \cite{wuc93}).
The two phase scenario in particular involves processes 
quite distinct from those involved in stellar formation.
Condensation of material is favored where and when proto-solar
disk temperatures are low enough to permit
condensation of volatiles.  This material in the
proto-solar disk nebula can then coalesce into planetesimals
which themselves can accrete into rock and volatile
bodies of substantial size (e.g.\ \cite{gw73}; \cite{ws93}; \cite{wc93}).
Once a rocky core of
10 - 30 $M_\oplus$ ($M_\oplus$ = $6\times10^{27}$\,g)
has formed, runaway accretion of the planet's gaseous envelope occurs
(\cite{lissauer}; \cite{bp86}; \cite{pol96}).
Recent Galileo results on the composition of Jupiter
have caused a downward revision in the estimate of
Jupiter's core mass from 10 - 30 $M_\oplus$ (\cite{hubbard84})
to the revised lower estimates of 0 - 12  $M_\oplus$
(Guillot, Gautier, \& Hubbard 1997).
Absence of a rocky core would support 
the direct condensation models, although even in these models, 
small cores of rock or ice may exist (\cite{boss97}).
A core mass of 12 $M_\oplus$ would be consistent
with either direct condensation or the core formation
and runaway accretion growth scenarios.

It is possible that the Galilean satellites formed directly
from a proto-Jovian gaseous accretion disk. 
If we make this assumption then the densities 
of these satellites suggest that pressures 
of about 1 bar ($10^6$ dynes cm$^{-2}$), required
for formation of methane and ammonia, were present
in the proto-Jovian disk (\cite{pri93}, \cite{pri81}).
This in turn suggests that the pressure in the proto-solar disk  
was probably far lower than that of the proto-Jovian disk.  
Lastly, protostars form from the collapse of a largely isotropic
molecular cloud, whereas planets form in proto-stellar disks.
Disks have thicknesses which may not be much more
than a planet's tidal or Hill sphere radius, and therefore
planet formation takes place in an environment which
is much closer to two-dimensional than it is to isotropic.

\subsection{The angular momentum problem}

Rotations observed in
molecular cloud cores present a problem in the collapse of
clouds, in that
angular momentum must be removed to collapse proto-stars.
The `angular momentum problem' for Jupiter is complicated
by the proto-solar disk because formation of the object
must involve the transfer of angular momentum to and from 
the proto-stellar disk and the proto-Jovian disk/planet system.   
This makes it difficult
to estimate a centrifugal radius for the Jovian proto-planet and
disk system.  Because of this difficulty 
we first consider the range
of possible values for the angular momentum of the proto-Jovian 
disk and planet system.  However if accretion into the 
Hill sphere occurs 
through the Lagrange points (of the planet/star system) 
then it is possible to estimate
a centrifugal radius and compare this radius to the radius of
the planet.  If this radius is large then it follows that
material accreted into the Hill sphere would likely go into an accretion
disk before falling onto the planet itself.

The planet core is likely to accrete material that falls within its
tidal or Hill sphere radius,
$r_H = D \left({M_p \over 3  M_\star}\right)^{1/3}$,
where $D$ is the semimajor axis of the planet's orbit, $M_\star$
is the mass of the sun (or central star) and $M_p$ is the mass of the planet.
This tidal radius therefore represents a maximum radius for
the proto-Jovian disk.
We can estimate this radius for the current Jupiter mass
and semi-major axis
\begin{equation}
{r_H } = 5.3 \times 10^{12} {\rm cm}
\left({D \over 5.2 {\rm AU} }\right) 
\left({M_p \over M_J}\right)^{1/3} 
\left({M_\star \over M_\odot}\right)^{-1/3} 
\end{equation}
where $M_J$ and $M_\odot$ are
the masses of Jupiter ($2\times10^{30}$ g) and
the sun ($2\times10^{33}$ g), respectively,
and see that even for a much smaller proto-Jovian planet mass the tidal radius
is a few hundred times larger than the current radius of
Jupiter ($7\times10^9$cm).

If accretion into the Hill sphere radius occurred through the
Lagrange points (at a distance of $r_H$ from the planet, and at
a low velocity in the rotating frame in which both
Jupiter and the sun are stationary), as seen in simulations (\cite{art96}),
then we can estimate the spin angular momentum of the accreting material.
The spin angular momentum per unit mass (with respect to Jupiter) would be 
$\sim r_H^2 \Omega_p$ (where 
$\Omega_p = \sqrt{GM_* \over D^3}$ refers 
to the orbital angular rotation rate of the planet about the star) 
resulting in rotation for the planet that is prograde with respect
to its orbit about the star (e.g.\ \cite{kuiper}).
If we estimate a centrifugal radius $r_c$ based on this spin angular momentum
we find that $r_c \sim  r_H/3$, still well outside the planet radius
even during the early formation stages of the collapsing planet.
It is therefore likely that 
material accreted into the Hill sphere from the
proto-stellar nebula would go into (and through) an accretion
disk before falling onto the planet itself.
The formation of giant gaseous palents therefore could be similar
to that of protostars in that some mechanism (such as an accretion disk
and accompanying outflow) is required to transfer angular momentum outwards
for accretion onto the central object (planet) to occur.

\section{Outflow Models}

\subsection{Accretion disk parameters}


We must first explore the physical conditions 
appropriate in a proto-Jovian accretion disk. 
If accretion from the proto-solar disk went into a proto-Jovian disk
before accreting onto the planet, then 
the accretion rate, $\dot M_D$, through the disk should be similar to that
estimated from non-disk formation models.
\cite{pol96_b} calculate an accretion rate of 
$10^{-6} M_\oplus$/yr (or $10^{-9} M_J$/yr; 
here $M_J$ is a Jupiter mass)  of planetesimals for the formation
of the rocky core, but after the planet reaches a critical
core mass, gas accretion occurs much faster, possibly greater than 
$10^{-3} M_\oplus$/yr (or $10^{-6} M_J$/yr).
This accretion rate is similar to the timescale estimated
for the dissipation of proto-stellar disks (e.g.\ \cite{skr90}) so we 
scale our relations from this accretion rate.

We must also estimate the magnetic field in the proto-Jovian disk.
There is evidence that there were strong magnetic fields
in the proto-solar disk from measurements of primitive meteorites
(see review in \cite{mor93}).  Inferred magnetic fields range
from 0.1 to 1 Gauss. Since these fields were the result of
fields averaged over a long time period, the meteorite remanence record
suggests that the presence of magnetizing fields as intense 
as a few Gauss existed in the proto-solar disk at a distance of several 
AU from the sun.  
If the proto-Jovian disk had a much higher pressure
($\sim 1$ bar) than that of the proto-solar nebula 
(\cite{pri93}, \cite{pri81}, \cite{lunine}) then 
the magnetic field therefore could be as high as $5 \times 10^3$ Gauss in the
proto-Jovian disk and remain within equipartition.
For comparison the current magnetic field of Jupiter is approximately 
10 Gauss but is generated from an internal dynamo.

Here we check that the above accretion rates $\dot M_D$ 
could be sustained and would be consistent with
the range of pressures and temperatures estimated for
a proto-Jovian accretion disk.  Accretion through the disk 
implies that there is dissipation of energy into the disk.  
For a self-luminous disk the temperature as a function of radius is
\begin{equation}
T(r)^4 = {3 G M_p \dot M_D \over 8 \pi \sigma_{SB} r^3}
\end{equation}
where $\sigma_{SB}$ is the Stefan-Boltzmann constant.  Using the accretion
rate mentioned above we estimate
\begin{equation}
T =  370 K
\left({ M_p \over M_J}\right)^{1/4}
\left({\dot M_D\over 10^{-6}M_J/{\rm yr}}\right)^{1/4}
\left({ r\over 10^{11} {\rm cm }}\right)^{-3/4}.
\end{equation}
The sound speed at this radius would be $\sim 2$ km/s implying
a reasonably thin disk $h/r \lesssim 0.2$ (from hydrostatic equilibrium,
where $h$ is the vertical scale height).
For a surface density $\Sigma = 10^4$ g cm$^{-2}$, the central disk pressure 
would be $\sim 5 \times 10^3$ dynes cm$^{-2}$ 
at this radius. This pressure is lower than
predicted based on the chemical composition (\cite{pri93}), but
still sufficiently high that a 10 Gauss field would still be
well within equipartition.  
For this surface density an $\alpha$ (accretion or viscosity) 
parameter of $\alpha \sim 10^{-3}$ -- $10^{-4}$ would
be consistent with the above accretion rate.
For a higher density disk (e.g.\ \cite{lunine}), 
a lower $\alpha$ would be consistent
with the above accretion rate.

Under conditions of low ionization fraction 
for an outflow to be launched the neutral component must
be well coupled with the ions (e.g.\ \cite{war93}).
This coupling is
described by $\eta$, the ratio of the dynamical time to the mean
collision time of a neutral atom in a dilute sea of ions 
(e.g.\ \cite{war93}; or equivalently \cite{shu94a});
$
\eta = {\gamma_i \rho_i r \over v_c},
$
where 
$v_c$ is the rotational velocity of a particle in a circular orbit.
When $\eta \gg 1$ the neutrals move with the ions and
so can be described as well coupled to the magnetic field.
Here 
$\gamma_i \sim 3 \times 10^{13}$ cm$^3$ g$^{-1}$ s$^{-1}$
is the drag coefficient between ions and neutrals
and $\rho_i$, the density of ions,
is the product of the atomic mass of the ions,
the ionization fraction and the total density.
We can then estimate
\begin{equation}
\eta \sim 1\times 10^8 \left({n_e/ n_H \over 10^{-12}}\right)
       \left({\rho \over 10^{-6} {\rm g~cm}^{-3}}\right)
       \left({M \over M_J}\right)^{-1/2}
       \left({r \over 10^{11} {\rm cm}}\right)^{3/2}
\end{equation}
For the disk with ionization state explored by \cite{takata_b}  
(with temperatures not largely different than in Eqn.~3) 
the density ranges
from $\rho = 10^{-3}$ to  $10^{-12}$ g cm$^{-3}$ for radii between 1 and
100 planet radii and at scale heights less than $z/h < 4$.
The ionization fraction 
(ranging from ${n_e\over n_H}  = 10^{-16}$ to $10^{-9}$)
is roughly inversely correlated with the 
density (compare Fig.~3 with Fig.~4 of \cite{takata}).
We conclude that over the whole model disk $\eta \gg 1$ and the
neutrals can be said to be well coupled to the ions.
However this does not mean that the particles are necessarily
well coupled to the magnetic field.

The magneto-centrifugal wind models can operate where the 
charged particles are sufficiently coupled to the field lines that they
are flung out centrifugally from the outer layers of the disk
(for a favorable field geometry).   This occurs when
the magnetic Reynold's number is high or $\gg 1$.
%
For the disk with ionization state explored by \cite{takata_b},
the magnetic Reynold's number
is only large ($> 1$) in the outer layers (greater than a 
few vertical scale heights
where ionization is produced mainly by galactic cosmic rays),
and within a few planetary radii of the planet (due to thermal ionization).
The magnetic field should primarily be important
in these regions.

\subsection{Magneto-centrifugally driven disk outflow models}

In the magneto-centrifugal disk wind or outflow models, 
the angular momentum of the wind is tied to that of the disk, 
and so the wind flux is directly related 
to the accretion rate, $\dot M_D$, through the disk.
A parameter $\epsilon$ (see \cite{pel92}, Eqn.~7.1)
\begin{equation}
\epsilon \equiv {\dot M_D     \sqrt{G M_p} \over 3 B^2 r^{5/2}}
\end{equation}
describes ${1\over 3}(r/r_A)^3$, the ratio of the radius $r$ 
to the Alfven radius, $r_A$,
of the field line passing through $r$.  
For a wind to be produced the field lines must be
more than $30^\circ$ from normal to the disk and $\epsilon < 1$.
Here $B$ is the magnetic field in the disk at $r$
and $G$ is the gravitational constant.

For these rough Jupiter sized parameters, assuming $r \sim 10^{11}$\,cm,
as a typical size for a circumplanetary disk (\cite{stevenson86}),
we have
\begin{equation}
\epsilon \sim 0.02
\left({\dot M_D     \over 10^{-6} M_J {\rm yr}^{-1} }\right)
\left({M_p    \over M_J         }\right)^{1/2}
\left({B      \over 10 {\rm Gauss}            }\right)^{-2}
\left({r      \over 10^{11} {\rm cm}    }\right)^{-5/2}.
\end{equation}
Since $\epsilon < 1$, 
we find that if the magnetic field in the disk is greater than 10 Gauss
and the inflow rate is not high then we would expect a wind or bipolar
outflow to be driven from the disk (assuming the field geometry is favorable).

For these models (\cite{pel92}) 
the mass flux $\dot M_w$ in the wind is related
to the accretion rate through the disk by
\begin{equation}
\dot M_w = f \dot M_D
\end{equation}
where 
\begin{equation}
f \approx \left({r \over r_A}\right)^2.
\end{equation}
The terminal speed of the outflow is 
\begin{equation}
v_\infty \sim v_c \left({r_A  \over r }\right)
\end{equation}
where $v_c$ is the velocity of rotation in the disk
at the radius from which the wind originates.

If the wind originates from the range of Jupiter's nearby
moons at $5 \times 10^{10}$ cm (the distance of Io from Jupiter) 
and $r_A/ r \sim 3$ (typical of solutions in \cite{pel92}) then 
the outflow velocity could be as large as $\sim 20$ km/s, 
and the mass outflow rate in the wind would be
$\sim 10^{-7} M_J$/yr (for an inflow rate of $10^{-6} M_J$/yr).

We note that a magneto-centrifugal disk wind would be 
launched at radii greater than a few planetary radii from the outer 
ionized layers of the disk.
It is unclear that there would be direct coupling
between accretion ($\dot M_D$) in a dense neutral disk 
and an outflow launched in the low density layers 
at a few scale heights above the disk
(where the ionization is sufficient that the MHD approximation is valid).
Since Eqn.~4 holds for lower accretion rates, 
if the outflow is only coupled to the outer layers of the disk,
then the mass outflow rate may be much lower than estimated above.

\subsection{Magneto-centrifugally driven stellar outflow models - The X-wind}

A more detailed alternative model is the magneto-centrifugally driven 
stellar and disk outflow model of \cite{shu94a_} and subsequent papers.
In this model the stellar magnetic field is sufficiently strong that
the disk is truncated at a radius $R_x$ given by 
\begin{equation}
R_x = \alpha_x \left({ \mu_p^4 \over G M_p \dot M_D^2}\right)^{1/7}
\end{equation}
(\cite{gho78}, \cite{gho79a}, \cite{gho79b})
where $\mu_p = B_p R_p^3$ is the magnetic moment of the planet, 
$R_p$ is the planet radius, 
and $B_p$ is the planet's bipolar magnetic field.
$\alpha_x$ is a dimensionless parameter which can be calculated
depending upon the details of the model and ranges from 
$0.5-1.2$ (\cite{shu97}).
\cite{shu94a_} argue that if the magnetic field is strong enough to truncate
the disk ($R_x/R_p > 1$), then it is automatically strong enough to launch a 
magneto-centrifugally driven wind.

For Jupiter sized parameters
\begin{equation}
{R_x \over R_p} =  1.3 \ \alpha_x 
\left({B_p \over 10 {\rm Gauss}}\right)^{4/7}
\left({R_p \over 2 \times 10^{10}{\rm cm}}\right)^{5/7}
\left({M_p \over M_J            }\right)^{-1/7}
\left({\dot M_D \over 10^{-7} M_J {\rm yr}^{-1}   }\right)^{-2/7}
\end{equation}
Here we have chosen to scale from a lower accretion 
rate more appropriate
for later stages of evolution and a smaller radius 
more appropriate for activity nearer the planet surface.
We see that if the inflow rates are moderate and the planetary magnetic field
is larger than 10 Gauss,
we find it is likely that the disk was truncated
by the magnetosphere of the planet, and so an X-wind 
could be driven from the planet+disk system.
Since the truncation is likely to occur within a few planetary radii,
we expect that the ionization state of the disk (due to thermal ionization) 
is sufficient to couple the magnetic field to the disk.

For the X-wind model $f\approx 1/3$
but this factor in proto-stars depends on the mass/radius relation
for deuterium burning.
For an accreting proto-Jovian planet the mass radius relation could
be substantially different (e.g.\ \cite{pol96},
\cite{guillot}, \cite{saumon})
and so 
we do not necessarily expect this factor to be similar.

For the X-wind  model the wind is most energetically
driven from the X-point or the radius at which
the disk rotates at the same speed as the planet. 
The speed of the outflow is then related to the velocity
of rotation at that radius as in Eqn.~9 above.
Given Jupiter's current angular rotation rate (of the magnetic field)
this radius is $1.6 \times 10^{10}$ cm and has 
a circular velocity of 28 km/s.  The proto-Jupiter 
would have been larger and rotating more slowly
making the radius of corotation somewhat larger. 
To estimate the wind speed we would then divide by $f$
which for moderate factors of $f\sim 1/3$ would imply
moderate outflow speeds of order $\sim 30$ km/s.

\subsection{Scaling from young stellar objects}

In this subsection we scale from the empirical relations
observed in proto-stellar outflows  
to estimate wind outflow rates and velocities.
This approach was introduced by \cite{wol90}.
The mechanical luminosity and force or thrust (rate of momentum
injection) of bipolar molecular outflows scales with the radiant
luminosity of the driving stellar source (\cite{lad85}, \cite{cab92}).
Observational estimates of the bolometric luminosity $L_{bol}$ of
the stellar source, $L_{CO}$ the outflow mechanical luminosity
and $F_{CO}$, the outflow momentum injection rate (force or thrust)
can be fit to the following scaling ``laws'' (\cite{cab92}, \cite{cab86}).
The subscript $CO$ here refers to the carbon monoxide line emission 
in which the proto-stellar 
outflows are studied.  We note that a proto-Jovian outflow
may not be bright in carbon monoxide emission for a variety of reasons
(e.g., Jupiter's composition), but consider these
laws for conceptual and scaling purposes.
The force law is the following
\begin{equation}
{F_{CO} c \over L_{bol}} \sim 2000 
  \left({L_{bol} \over L_\odot}\right)^{-0.3}
\end{equation}
where $c$ is the speed of light, 
$F_{CO} \sim \dot M_w v_w$ and $v_w$ is a wind velocity
(though there is an added complication that much of the material
observed in the outflow could be entrained gas),
and $L_\odot$ is the solar luminosity, $3.9\times10^{33}$ergs/sec.
The luminosity law:
\begin{equation}
{L_{CO} \over L_{bol}} \sim 0.04 
  \left({L_{bol}\over L_\odot}\right)^{-0.2}
\end{equation}
where  $L_{CO} = \dot M_w v_w^2$.

Jupiter during its formation at early times has 
$L_{bol} \sim 10^{-6}L_\odot$ at a few times $10^6$ years after formation
(\cite{bur97}) 
so if the physical processes for the early accretion phase are similar
to those in proto-stars then we would predict that 
\begin{equation}
F_{CO} \sim 2.6 \times 10^{-6} 
\left({M_J \over {\rm yr}}\right) 
\left({{\rm km} \over {\rm s}}\right) 
\left({L_{bol} \over 10^{-6} L_\odot}\right)^{0.7}
\end{equation}
and 
\begin{equation}
L_{CO} \sim 4.0 \times 10^{-6}
\left({M_J \over {\rm yr}}\right) 
\left({{\rm km} \over {\rm s}}\right)^2
\left({L_{bol} \over 10^{-6} L_\odot}\right)^{0.8}
\end{equation}
These scaling laws suggest that outflows of $10^{-7} M_J$/yr
could occur for moderate outflow velocities 
of order 10 km/s; the same sizes 
as we estimated could be possible in the theoretical
models discussed above.

\section {Summary and Discussion}
In this paper we have considered the possibility that the formation
of Jovian planets involved an outflow/accretion phase
similar to that observed in proto-stars.
There are a number of similarities in the expected formation of the
these objects but also a number of significant differences.
We assume here that accretion onto the planet took place through a disk.
Since the Hill sphere (tidal) radius of Jupiter is a few
hundred times larger
than its current radius, it is unlikely that gas and planetesimals
captured by the gravitational field of Jupiter had little spin angular
momentum.  If accretion into the Hill sphere occurs 
through the Lagrange points then 
the accreting material should have spin angular momentum
(with respect to the planet) equivalent to a centrifugal
radius of 1/3 of the Hill sphere radius, again much larger than the planet
radius.  This suggests that formation of Jupiter could
have involved accretion of a significant fraction
of the planet mass through a disk.

Order of magnitude estimates of magneto-centrifugal disk wind
and stellar + disk (X-wind) models suggest that if the 
accretion of the planet occurs through a disk,
at a rate that is not extremely high (or less than $10^{-6} M_J$/yr), 
and if the magnetic field in the disk or planet was larger
than 10 Gauss, an outflow could be driven from a proto-Jovian
planet.  Here we have assumed disk accretion rates 
appropriate for formation of Jupiter on timescales of $10^6-10^7$
years.  However winds could also be driven at lower disk accretion rates
with lower magnetic fields from a lower mass disk
that might have existed at a later stage of formation or was
formed subsequent to formation of the planet (e.g.\ \cite{kor91}).
The feasibility of an outflow depends on the magnetic field
geometry and the ionization structure of the disk.
While the X-wind models are launched at small radii where 
the ionization state is sufficiently large to couple
the magnetic field to the gas, the disk wind models
can only be launched from the outer diffuse layers of the disk.
Future work should therefore explore self consistent models for such 
an accretion disk to determine the feasibility of outflows
at varying accretion rates.

If Jovian planets indeed form via an accretion disk+outflow
process similar to that in proto-stars then there are number of possible
interesting consequences which we list below.

1) The resulting planet size (and possibly percentage of gas mass) 
could be determined by the efficiency of the
outflow rather than by the gas supply or dissipation of
the proto-solar disk (as suggested for proto-stars;
\cite{shu87} and \cite{nak95}).

2) It is possible that some fraction of mass similar to the final
planet mass could have been processed chemically 
through proto-Jovian disks during the time of planet formation.
This gas mass would be rich in molecules such
as methane and possibly ammonia which require pressures
higher than typical of the proto-solar disk but which could
have existed in the proto-Jovian disk (\cite{pri93}, \cite{pri81}).
Some of this gas ejected as part of the outflow could 
become part of the proto-solar
disk or part of the Oort cloud 
leaving a chemical signature of planetary processing 
which might be detected in later evolutionary 
stages of proto-stellar
disks or perhaps in the solar neighborhood in terms
of abundance variations in meteorites or comets.
Gas ejected by an outflow from a Jupiter-sized planet could
escape the solar system altogether, whereas outflows
from lower mass planets might be more likely to enrich the outer solar
system.  Chemical variations in the different circumplanet disks
could lead to quite a wide variety of molecules dispersed.
This dispersion could be compared to the efficiency
of various mixing processes in the proto-solar disk.

An outflow could also lead to variations in abundances
in the forming planet.  
This might lead to alternative explanations
for Jupiter's near solar abundance, water content, 
high carbon and sulfur content
and depleted oxygen content (\cite{niemann};
as opposed to chemical enrichment by infalling planetesimals 
subsequent to formation).

3) The rotation of Jovian planets could be
interpreted naturally as resulting from planet/disk interactions
(\cite{takata})
rather than from three-body or hydrodynamic affects
as gas flowed into the Hill sphere (as in \cite{coradini}; 
\cite{art96}) or from angular momentum transfer between 
the core and envelope (\cite{kor91}).

4) Outflows from proto-Jovian planets and their accretion disks 
could provide a way to detect forming planets.
These outflows could involve dispersal of a significant fraction
of the final planet mass into the proto-stellar disk or even
out of the stellar gravitational field (or solar system) altogether.
This dispersed material might be optically thin (making
it easier to detect) and would have a chemical 
and kinematic signature differing from the proto-stellar disk.
The inner regions of the circum-planet 
accretion disks would be significantly hotter than the surrounding
proto-stellar nebula and with luminosities similar 
to the young planet itself, but covering
a broader range of wavelengths.
Some preliminary prospects for detection are discussed in the appendix.

\acknowledgments

We thank the referee for many comments which have improved this paper.
We also acknowledge support from NSF grant AST-9529190 to M.~and G.~Rieke
and NASA project no. NAG-53359. D.~T.\ is supported by
an NSF Graduate Research Fellowship.
This work could not have been carried out without
helpful discussions and correspondence with
J.~Lunine,  J.~Najita, C.~Strom, J.~Valenti, F.~Melia,
A.~Nelson, D.~DeYoung, H.~Chen, W.~Hubbard, M.~Hanson,
L.~Golub, S.~Saar, A.~Dessler, E.~Deluca, T.~Ruzmaikina,
G.~Rieke, E.~Young, D.~Sudarsky, M.~Rieke, D.~Willner, J.~Bechtold,
J.~Beijing, M.~Meyer, A.~Sprague, R.~Brown, T.~Fleming and M.~Sykes.

\newpage
\appendix
\section{Some Prospects For Detection of Forming Jovian Planets}

The circum-planet accretion disk itself could have a luminosity
(reprocessed radiation from the planet or/and from accretion)
$L_\nu = \nu  F_\nu$ similar to the bolometric luminosity
of the planet.   However its spectral energy distribution
would be such that it would be bright over a larger range 
of wavelengths than the planet (e.g.\ \cite{ada87})
which should be closer to a black-body.
This implies that it may be possible to search for longer wavelength
(e.g.\ mid-IR) emission from the hot regions of 
circum-planet accretion disks at radii from the central star
that are not expected to be bright at these wavelengths.
A proto-Jovian accretion disk could have a flux at $\lambda = 10 \mu$m of
$$
F_\nu \sim 1 {\rm mJy} 
\left({L_{bol}  \over 10^{-6} L_\odot} \right)  
\left({\lambda  \over 10 \mu {\rm m}}  \right)   
\left({D        \over 10 {\rm pc}   }  \right)^{-2}
$$
which is detectable by ISO, and SIRTF and other mid-IR cameras.  
However high angular
resolution would be required to differentiate this emission
from the proto-stellar disk emission.

To explore the possibility of detection of proto-Jovian
outflows we first estimate the column density of material
that could have been ejected.
If  1\% of a Jupiter mass is ejected over a circular region of radius
5AU then with a Jupiter abundance of ammonia or methane 
($\sim 2 \times 10^{-3}$ number density
with respect to H$_2$; \cite{niemann})
we estimate a column depth in methane or ammonia of
$1.3 \times 10^{21}$ cm$^{-2}$ and a column of hydrogen atoms 
of $N_H \sim 6 \times 10^{23}$ atoms cm$^{-2}$ 
(which could be in molecular form).
The cooling time is short so much of this gas could be cold,
although ambipolar diffusion may result in heating the outflowing
gas near the planet (as is predicted for proto-stellar outflows;
Safier 1993a,b).   The level of column depth suggests that extinction 
against the central source or planet could be high 
for typical galactic dust to gas ratios.

We now consider some highly sensitive ways for detecting cold
gas in emission.  Cold ($\sim 10 K$) ammonia can be detected 
at the mJy level as emission in the inversion
transitions (23 GHz) at a column of 
$N_{\rm NH_3} \sim 10^{14}$ cm$^{-2}$ (e.g.\ \cite{gom94}),
so although we expect our emitting region to be small,
(and so a factor dependent on the beam size will affect
the detection limit), it is possible
that ammonia in young stellar systems could be detected in the
vicinity of nearby young stars during the time of planet formation.
Ammonia is also enhanced in proto-stellar outflows 
(e.g.\ \cite{taf95}).

Carbon monoxide can currently be detected at the 10 mK level 
in emission in the rotational transitions 
(e.g.\ $^{12}$C$^{16}$O(1--0) or (2--1)) 
in the millimeter and submillimeter at a column depth
of $N_{\rm CO} \sim 10^{16}$ cm$^{-2}$.  Detection limits in carbon
monoxide emission for nearby young stars could
therefore place limits on the fraction of residual cold gas phase 
carbon monoxide in young solar systems (e.g.\ \cite{koe95}).
However detection of a Jovian outflow might easily
be confused with other phenomena 
such as an outflow from the central star or inflow
from infalling molecular clumps into the whole stellar and disk system.
A combination of high spectral resolution and a variety
of molecular tracers (e.g.\ \cite{bac97})  
would be required to differentiate between phenomena.

Ejected outflow material could also be detected in absorption
against the central source.
Molecules such as water, methane, and carbon dioxide
in the form of ices with column depths of $10^{16}  - 10^{17}$ cm$^{-2}$
(and also silicates at 10$\mu$m) have been detected in absorption
against 100 Jy sources with ISO (e.g.\ \cite{whi96}).
Absorption against a central star is likely if an extended (out
of the ecliptic) and a high column density of these molecules exists.
For comparison, a solar type star at 100 pc is about 100mJy at 10$\mu$m.
This suggests that the next wave of near-IR spectrographs (e.g.\ SIRTF)
could observe young solar-type main sequence stars out to a few
hundred pc and search
for absorption of ices and silicates against the central star in the
mid-infrared ($2-30\mu$m).  High spectral resolution observations
could also be done of molecular absorption lines at optical wavelengths.

Outflow velocities of order 20 km/s are large enough
to drive shocks that would
emit at $\sim 2000 K$.  This opens up the possibility of detecting shocked
molecular hydrogen and other tracers of hot dense
gas.  Outflowing gas near the planet could also
be detected in emission at relatively hot temperatures of $100-1000 K$
(typical of the forming disk and planet)
in the near and mid-IR lines.  At these wavelengths dilution caused by
the central stellar source is likely to be a problem, however
the lines emitted would not be characteristic of emission
from the central star and the kinematics of the emitting gas would
be peculiar: that of an outflow coupled with the orbital
motion of the planet.

Possible targets for a search for forming proto-Jovian planets
would be $\sim 10^6$ year old
stars lacking evidence for energetic accretion. 
These stars would be likely
sites for currently forming Jovian planets.
Candidates would be X-ray bright, T Tauri stars or 
solar-type stars in relatively (but not extremely) 
young stellar clusters.

\clearpage



\begin{thebibliography}{}

\bibitem[Adams et al.~1987]{ada87}
Adams, F.~C., Lada, C.~J., \& Shu, F.\ 1987, ApJ, 312, 788


\bibitem[Anderson et al.~1997]{anderson}
Anderson, J.~D., Lau, E.~L., Sjogren, W.~L., Schubert, G., \&
Moore, W.~B.~1997, Nature, 387, 264

\bibitem[Andre, Ward-Thompson \& Barsony 1993]{and93}
Andre, P., Ward-Thompson, D., \& Barsony, M.~1993, ApJ, 406, 122

\bibitem[Artimowicz \& Lubow 1996]{art96}
Artimowicz,  P., \& Lubow, S.~H.~1996, ApJ, 467, L77

\bibitem[Bachiller 1997]{bac97}
Bachiller, R.~1997,
in "Molecules in Astrophysics: Probes and Processes, IAU Symposium No. 178,
ed.~E.~W.~van Dishoeck, Kluwer Academic Publishers, Dordrecht, 103

\bibitem[Bally \& Lane 1991]{bal91}
Bally, J., \& Lane, A.~P.~1991, in The physics of star formation and early
stellar evolution, eds.~C.~J. Lada and N.~D. Kylafis, Kluwer Academic Publishers,
Dordrecht, p.~471

\bibitem[Blandford \& Payne (1982)]{bla82}
Blandford, R.~D., \& Payne, D.~G.~1982, MNRAS, 199, 883

\bibitem[Bodenheimer \& Pollack 1986]{bp86}
Bodenheimer, P. \& Pollack, J.~B.~1986, Icarus, 67, 391

\bibitem[Bontemps et al.~1996]{bon96}
Bontemps, S., Andre, P., Terebey, S., \& Cabrit, S.~1996, A\& A, 311, 858

\bibitem[Boss 1997]{boss97}
Boss, A.~P. 1997, Science, 276, 1836.

\bibitem[Brown et al.~1997]{rhb}
Brown, R.~H., Cruikshank, D.~P., Pendleton, Y., Veeder,
	G.~J.~1997, Science, 276, 937

\bibitem[Burrows et al.~1997]{bur97}
Burrows, A., Marley, M., Hubbard, W.~B., Lunine, J.~I., Guillot, T., 
Saumon, D., Freedman, R., Sudarsky, D., \& Sharp, C.~1997, 
ApJ,  in press

\bibitem[Cabrit \& Bertout 1992]{cab92}
Cabrit, S., \& Bertout, C.~1992, A\&A, 261, 274

\bibitem[Cabrit \& Bertout 1986]{cab86}
Cabrit, S., \& Bertout , C.~1986 , ApJ, 307, 313

\bibitem[Coradini et al.~1989]{coradini}
Coradini, A., Cerroni, P., Magni, G., \& Federico, C.~1989, in
Origin and Evolution of Planetary and Satellite Atmospheres,
eds.~S.~K. Atreya, J.~B. Pollack, \& M.~S. Matthews,
The University of Arizona Press, Tucson, p.~723

\bibitem[Gammie 1996]{gam96}
Gammie, C.~F.~1996, ApJ, 457, 355

\bibitem[Ghosh \& Lamb 1978]{gho78}
Ghosh, P. \& Lamb, F.~K.~1978, ApJ, 223, L83

\bibitem[Ghosh \& Lamb 1979a]{gho79a}
Ghosh, P. \& Lamb, F.~K.~1979a, ApJ, 232, 259

\bibitem[Ghosh \& Lamb 1979b]{gho79b}
Ghosh, P. \& Lamb, F.~K.~1979b, ApJ, 234, 296

\bibitem[Glassgold, Najita \& Igea 1997]{gla97}
Glassgold, A.~E., Najita, J., \& Igea, J.~1997, ApJ, 480, 344

\bibitem[Goldreich \& Ward 1973]{gw73}
Goldreich, P., \& Ward, W.~R.~1973, ApJ, 183, 1051

\bibitem[Gomez et al.~1994]{gom94}
Gomez, J.~F., Curiel, S., Torrelles, J.~M., Rodr\'iguez, L.~F.,
Anglada, G., \& Girart, J.~M.~1994, ApJ, 436, 749

\bibitem[Guillot et al.~1996]{guillot}
Guillot, T., Burrows, A., Hubbard, W.~B., 
Lunine, J.~I., \& Saumon, D.~1996, \apjl, 459, L35

\bibitem[Guillot et al.~1997]{guillot97}
Guillot, T., Gautier, D., Hubbard, W.~B.~1997, Icarus, 130, 534

\bibitem[Hartmann \& MacGregor 1982]{har82}
Hartmann, L., \& MacGregor, K.~B.~1982, ApJ, 259, 180

\bibitem[Heyvaerts \& Norman 1989]{hey89}
Heyvaerts, J.,\& Norman, C.~1989, ApJ, 347, 1055

\bibitem[Hubbard 1984]{hubbard84} 
Hubbard, W.~B.~1984, Planetary Interiors,
(New York: Van Nostrand-Reinhold)

\bibitem[Isbell, Dessler \& Waite 1984]{isb84}
Isbell, J., Dessler, A.~J., \& Waite, Jr., J.~H.~1984,
Journal of Geophysical Research, 89, 10716

\bibitem[Koerner \& Sargent 1995]{koe95}
Koerner, D.~W., \& Sargent, A.~I.\ 1995, AJ, 109, 2138

\bibitem[Ko\"nigl 1989]{kon89}
Ko\"nigl, A.~1989, ApJ, 342, 208

\bibitem[Ko\"nigl 1991]{kon91}
Ko\"nigl, A.~1991, ApJ, 370, L39

\bibitem[Korycansky et al.~1990]{kor90}
Korycansky, D.~G., Bodenheimer, P., Cassen, P. \& Pollack, J.~B.~1990,
Icarus, 84, 528

\bibitem[Korycansky et al.~1991]{kor91}
Korycansky, D.~G., Bodenheimer, P., \& Pollack, J.~B.~1991,
Icarus, 92, 234

\bibitem[Kuiper 1951]{kuiper}
Kuiper, G.~P., in "Astrophysics", ed.~J.~A. Hynek, 1951, 
McGraw-Hill Book Co, Inc., New York, 357 

\bibitem[Lada 1985]{lad85}
Lada, C.~J.~1985, ARAA, 23, 267 

\bibitem[Levy et al.~1991]{lev91}
Levy, E.~H., Ruzmaikin, A.~A., \& Ruzmaikina, T.~V.~1991, 
in `The Sun in Time', 
eds.~C.~P. Sonnett and M.~S. Giampapa and M.~S. Maththews ,
Tucson, University of Arizona Press, p.~589 

\bibitem[Lissauer 1993]{lissauer}
Lissauer, J.~J.~1993, ARAA, 31, 129

\bibitem[Lunine \& Stevenson 1982]{lunine}
Lunine, J.~I., \& Stevenson, D.~J.~1982, Icarus, 52, 14

\bibitem[McKinnon 1997]{mck97}
McKinnon, W.~B.\ 1997, Icarus, 130, 540

\bibitem[Mestel (1968)]{mes68}
Mestel, L.~1968, MNRAS,  138, 359

\bibitem[Morfill, Spruit \& Levy 1993]{mor93}
Morfill, G., Spruit, H. \& Levy, E.~H.~1993, 
in Protostars and Planets III, eds.~E.~H.~Levy, \& J.~I.~Lunine,
The University of Arizona Press, Tucson, p. 939


\bibitem[Morgan \& Bally 1991]{mor91}
Morgan, J.~A., \&  Bally, K.~1991, ApJ, 372, 505

\bibitem[Nakano, Hasegawa \& Norman 1995]{nak95}
Nakano, T., Hasegawa, T., \& Norman, C.~1995, ApJ, 450, 183

\bibitem[Niemann et al.~1996]{niemann}
Niemann, H.~B.~et al.~1996, Science, 272, 846


\bibitem[Pollack, Lunine \& Tittemore 1990]{pol90}
Pollack, J.~B., Lunine, J.~I., \& Tittemore, W.~C.\
1990, in Uranus, ed.~Bergstrahl, J.~T., Miner, E.~D.,
\& Matthews, M.~S.\ (Tucson: University of Arizona
Press), 469


\bibitem[Pelletier \& Pudritz 1992]{pel92}
Pelletier, G. \& Pudritz, R.~E.~1992, ApJ, 394, 117

\bibitem[Podolak et al.~1993]{pod93}
Podolak, M., Hubbard, W.~B. \& Pollack, J.~B.~1993, 
in Protostars and Planets III, eds.~E.~H.~Levy, \& J.~I.~Lunine,
The University of Arizona Press, Tucson, p. 1109

\bibitem[Pollack et al.~1996]{pol96}
Pollack, J.~B., Hubicky, O., Bodenheimer, P., Lissauer, J.~J., Podolak, M., \&
Greenzwig, Y.~1996, Icarus, 124, 62

\bibitem[Prinn 1993]{pri93}
Prinn, R.~G.~1993, 
in Protostars and Planets III, eds.~E.~H.~Levy, \& J.~I.~Lunine,
The University of Arizona Press, Tucson, p.~1005

\bibitem[Prinn \& Fegley 1981]{pri81}
Prinn, R.~G., \& Fegley, B.~1981, ApJ, 249, 308

\bibitem[Pudritz \& Normal 1983]{pud83}
Pudritz, R.~E., \& Norman, C.~A.~1983, ApJ, 274, 677

\bibitem[Safier 1993a]{saf93a}
Safier, P.~N.\ 1993a, ApJ, 408, 115

\bibitem[Safier 1993b]{saf93b}
Safier, P.~N.\ 1993b, ApJ, 408, 148

\bibitem[Saumon et al.~1996]{saumon}
Saumon, D., Hubbard, W.~B., Burrows, A., Guillot, T.,
	Lunine J.~I., \& Chabrier G.~1996, \apj, 460, 993

\bibitem[Shu, Adams \&  Lizano 1987]{shu87}
Shu, F.~H., Adams, F.~C., \& Lizano, S.~1987, ARAA, 25, 23

\bibitem[Shu et al.~(1988)]{shu88}
Shu, F.~H., Lizano, S., Ruden, S., \& Najita, J.~1988, ApJ, 328, L19

\bibitem[Shu 1993]{shu93}
Shu, F.~H.~1993, in Molecular Clouds and Star Formation, eds.~C.~Yuan \& J.~You 
World Scientific, Singapore, p.~97

\bibitem[Shu et al.~1994a]{shu94a}
Shu, F., Najita, J., Ostriker, E., Wilkin, F., Ruden, S.
\& Lizano, S.~1994a, ApJ, 429, 781

\bibitem[Shu et al.~1994b]{shu94b}
Shu, F.~H., Najita, J., Ruden, S.,  \& Lizano, S.~1994b, 
ApJ, 429, 797

\bibitem[Shu et al.~1995]{shu95}
Shu, F.~H., Najita, J., Ostriker, E., \& Shang, H.~1995, 
ApJ, 455, L155

\bibitem[Shu \& Shang 1997]{shu97}
Shu, F.~H., \& Shang, H.~1997,  in Herbig-Haro Flows and the 
Birth of Low Mass Stars, IAU Symp. 182, eds.~B.~Reipurth and
C.~Bertout, Kluwer Academic Publishers, Dordrecht, p.~225

\bibitem[Skrutskie et al.~1990]{skr90}
Skrutskie, M.~F., Dutkevitch, D., Strom, S.~E., Edwards, S.,
Strom, K.~M., \& Shure, M.~A.~1990, AJ, 99, 1187

\bibitem[Stevenson et al.~1986]{stevenson86}
Stevenson, D.~J., Harris, A.~W., \& Lunine, J.~I.\ 1986, 
in Satellites, eds.~J.~A.~Burns, \& M.~S.~Matthews,
The University of Arizona Press, Tucson, p.~39

\bibitem[Tafalla \& Bachiller 1995]{taf95}
Tafalla, M., \& Bachiller, R.\ 1995, ApJ, 443, L37

\bibitem[Takata \& Stevenson 1996]{takata}
Takata, T., \& Stevenson, D.~J.~1996, Icarus, 123, 404

\bibitem[von Zahn \& Hunten 1996]{vzh96}
von Zahn, U., \& Hunten, D.~M.~1996, Science, 272, 849

\bibitem[Wardle \& Ko\"nigl 1993]{war93}
Wardle, M., \& Ko\"nigl, A.~1993,  ApJ, 410, 218

\bibitem[Weidenschilling \& Cuzzi 1993]{wc93}
Weidenschilling, S.~J., \& Cuzzi, J.~N.\
1993, in Protostars and Planets III, eds.~E.~H.~Levy, \& J.~I.~Lunine,
The University of Arizona Press, Tucson, p.~1031

\bibitem[Wetherill \& Stewart 1993]{ws93}
Wetherill, G.~W., \& Stewart, G.~R.~1993, Icarus, 106, 190

\bibitem[Whittet et al.~1996]{whi96}
Whittet, D.~C.~B., et al.~1996, A\&A, 315, L357

\bibitem[Wolk \& Beck (1990)]{wol90}
Wolk, S~.J., \& Beck, S.~C.~1990, PASP, 102, 745

\bibitem[Wuchterl 1993]{wuc93}
Wuchterl, G.~1993, Icarus, 106, 323 

\bibitem[Pelletier \& Pudritz (1992)]{pel92b}
\bibitem[Shu et al.~(1994a)]{shu94a_}
\bibitem[Pollack et al.~(1996)]{pol96_b}
\bibitem[Takata \& Stevenson (1996)]{takata_b}

\end{thebibliography}
\end{document}